\documentclass[a4paper, 11pt]{article}
\pdfoutput = 1
\usepackage{jcappub}
\usepackage[T1]{fontenc}
\usepackage{hyperref}
\usepackage{appendix}
\usepackage{color,soul}
\usepackage{subfigure}

\title{Variability in Quasar Light Curves: using quasars as standard candles}
\author[1]{R. Solomon,\note{Corresponding author.}}
\author{and D. Stojkovic,}

\affiliation{HEPCOS, Department  of  Physics,  SUNY  at  Buffalo,  \\239 Fronczak Hall, Buffalo,  NY  14260-1500, USA}

\emailAdd{rancesol@buffalo.edu}
\emailAdd{ds77@buffalo.edu}

\abstract{
A relation between the variational slope, $s_F$, and the mean absolute magnitude, $\langle M \rangle$, in the light curves of 58 spectroscopically confirmed quasars is measured with a dispersion of 0.15dex.
Assuming it holds for quasars in general, not only does this relation add to our working knowledge of quasar variability but it also shows great promise at accurately measuring luminosity distance to a quasar in a model independent way.
An accurate, model independent measure of the luminosity distance would allow quasars to be added to the cosmic distance ladder, easily extending the ladder out far beyond the redshifts accessible to type Ia supernovae where cosmological parameters can be better constrained.
}

\begin{document}
\maketitle
\flushbottom

%\\\\\\\\\\\\\\\\\\\\\\\\\\\\\\\\\\\\\\\\\\\\\\\\\\\\\\\\\\\\\\\\\\\\\\\\\\\\\
%/////////////////////////////////////////////////////////////////////////////

\section{Introduction}
\label{sec:intro}

The $H_0$-tension is at the forefront of many cosmological discussions today.
Since the mere $\approx 8.5\%$ difference between the early and late time $H_0$ measurements is sitting at a $4 - 6\sigma$ discrepancy we could be looking at either new physics or systematic errors in multiple observations.
We do not know of course which it is, but, in either case, the discrepancy is of great interest to the community and its study has instigated many new observational methods and theoretical possibilities.
In this article we focus on a new late time observational method.

The main approach to late time measurements is by building up the cosmic distance ladder in a similar manner to Hubble's original work in his measurement of the expansion rate.
At present there are three common rungs on the distance ladder: geometrical distance measurements, variable stars, and type Ia supernovae (SNe).
Each rung is anchored by the previous with SNe measuring out at the farthest redshifts of $z\approx0.15$ as of the $SH0ES$ 2016 release, ref.~\cite{Riess:2016jrr}.\footnote{An additional nine SNe in the range $1.5 < z < 2.3$ has been used in ref.~\cite{Riess:2017lxs} in order to probe the expansion rate towards the end of matter domination when dark energy is beginning to contribute significantly.}
The distances to SNe combined with the recessional velocity inferred by their redshift gives a measure of the expansion rate throughout our recent history.
And although local measurements are not able to directly constrain H$_0$ they are able to directly constrain $\Omega_m$, today's fractional matter density, thereby reducing the shared parameter space with H$_0$.

To improve the accuracy of the distance ladder one could add more anchors, add more SNe at greater redshifts, or add a new rung of astronomical objects at greater redshifts.
For this last reason there have been many attempts at incorporating quasars as they are among the most distant observable objects.
But much is still unknown about a quasar's intrinsic properties, making their highly irregular behaviors troublesome for phenomenological uses.
Currently the most promising attempt is being made in refs.~\cite{Risaliti:2016nqt,Lusso:2017hgz,Salvestrini:2019thn,Lusso:2020pdb} through a nonlinear relation between the X-ray and UV luminosities.
Although it had been previously known, the $L_X - L_{UV}$ relation had too large of dispersion ($\sim$0.35dex) to make accurate predictions of the luminosity distance.
However the authors were able to argue that much of the previously observed dispersion was not intrinsic to the quasars and could be subtracted out by choosing clean data sets, thereby reducing the dispersion to 0.2dex. 
Their results show that quasars prefer a lower luminosity distance at high redshifts than that predicted by $\Lambda$CDM (see Figure 9 of ref.~\cite{Lusso:2020pdb}).
The authors proceed to show that the lower luminosity distance at high redshifts can be better described with a time varying dark energy equation of state but it has been pointed out elsewhere (ref.~\cite{Banerjee:2020bjq}) that this part of the analysis may have a systematic error due to the choice of fitting function.
In ref.~\cite{Luongo:2021nqh} the authors discuss the possibility that the lower luminosity distances could point instead towards a violation of an FLRW metric due to an alignment of deviations from $\Lambda$CDM along the axis of the cosmic dipole moment in excess of that predicted.

The present article too is concerned with a phenomenological relation shared between quasars.
The method we discuss focuses on the measurement of the variational rate, $s_F$, of the absolute light curves (i.e. the slopes of small time scale variations in the rest frame flux of the quasar).
We find that $s_F$ shares a tight correlation with the mean absolute magnitude, $\langle M \rangle$, with a strikingly low dispersion of 0.15dex for 58 spectroscopically confirmed quasars from the MACHO dataset.
Furthermore, this dispersion results from a minimal selection criteria of the quasar light curves.
By assuming this relation holds for quasars in general, we show that one can obtain a quasar's luminosity distance from the measured apparent light curves.
This method originates in principle from ref.~\cite{Dai:2012wp} which had attempted to use quasars as standard clocks.

We address the details of the data used for the study in section~\ref{sec:data} while section~\ref{sec:findingmethod} discusses the finding methods used to measure the variational rate which we apply to 58 quasars from the MACHO survey.
Section~\ref{sec:dl} derives the luminosity distance from the measured apparent fluxes and contains a demonstration using 304 poorly sampled quasars from the Sloan Digital Sky Survey (SDSS).
Section~\ref{sec:drw} applies our analysis to light curves modeled by a damped random walk and discusses agreements with previous findings.
We finish with conclusions and a brief discussion of future work in section~\ref{sec:conclusion}.

%\\\\\\\\\\\\\\\\\\\\\\\\\\\\\\\\\\\\\\\\\\\\\\\\\\\\\\\\\\\\\\\\\\\\\\\\\\\\\
%/////////////////////////////////////////////////////////////////////////////

\section{The Quasar Samples}\label{sec:data}
For reasons which will be discussed in section~\ref{sec:findingmethod}, we require the quasar light curves to be sampled with a relatively regular and high density.
More than one observation per month is optimal for the analysis but sufficient results can still be obtained with less frequent observations.
This requirement reduces the number of previously available light curves that can be used in our current analysis since most prior quasar studies sample much less regularly.
However, the regularly sampled light curves from microlensing projects (MACHO, OGLE, etc.) and strong lensing time delay projects (COSMOGRAIL) can in principle be repurposed for our analysis.

Thanks to the efforts of Geha et. al. in ref.~\cite{MACHO:2002cer} 59 quasars from the MACHO project have been confirmed and made available for public use.\footnote{\url{http://www.astro.yale.edu/mgeha/MACHO/}}
Assuming the microlensing events are infrequent enough then one can easily avoid them or subtract them out of the light curves.
None of the 59 light curves show obvious signs of microlensing events, but quasar 42.860.123 has a significantly low sampling and will be neglected in this work.
The OGLE project on the other hand has observed over 700 quasars (see ref.~\cite{OGLE:2013bhd}) -- the light curves of which are expected to be in the coming OGLE IV release but are not yet available at the writing of this article.

The multiply lensed quasars from the COSMOGRAIL collaboration have the potential to be highly beneficial to our analysis since their time delays effectively extend a quasar's observation time which gives better statistics to our analysis.
Also, the overlapping regions between two matched light curves could reduce the uncertainty in the quasar's intrinsic variations.
Having said that, the purposes of the time delay measurements are concerned only with matching the light curves from one quasar at a time and do not require a universal calibration among different quasars.
The calibrations are instead done with respect to stable stars within a small viewing angle of each strongly lensed quasar thereby making their apparent magnitudes shifted by some undetermined value unique to each quasar.
One could in principle recalibrate these light curves using the calibrations given by the COSMOGRAIL team; however, the additional effect of the lens is not very well understood and can cause an uncertain amount of brightening in the individual images.
These effects would require further study before the light curves can be significantly considered towards our purposes.

Focusing now on the 58 remaining quasars from the MACHO project, the apparent magnitudes are supplied in both the v- and r-bands.
We make the usual conversion to absolute magnitudes,
\begin{equation}\label{eq:abs_M}
    M = m - 5(\log D_L-1) - K(z)
\end{equation}
where $m$ and $M$ are the apparent and absolute magnitudes, respectively, and $D_L$ is the luminosity distance.
The K-correction, $K(z)$, is an approximate correction for the apparent shift of the spectral energy distribution across the narrow band filters used in observations.
For example, if the quasar's flux varies over its spectrum as $f\sim\nu^\delta$ then two identical quasars at different redshifts (but measured with the same filter) would display different absolute magnitudes.
In magnitude form the K-correction goes as
\begin{equation}
    K(z) = -2.5(1+\delta)\log(1+z)
\end{equation}
with the canonical spectral index $\delta = -0.5$ being taken (see ref.~\cite{Kcorrection:2000}).
For proof of concept $D_L$ has been calculated from the redshift of each quasar using a standard cosmology with $\Omega_m = 0.3$, $\Omega_{\Lambda} = 0.7$, and $h=0.7$.
This makes our present work model dependent, but using a larger collection of quasars at small redshift or ones with relative distances determined through cross-calibrations with SNe would allow us to avoid this model dependence in future work.

Greater discrimination in the variational slope can be found using the flux instead of magnitude measurements so we define the absolute flux as
\begin{equation}
    F = F_0 10^{-M/2.5}
\end{equation}
where $F_0 = 0.227meV/m^2/sec/hz$ is a v-band specific calibration constant which we set to unity ($F_0 \equiv 1$).
For the remainder of the article, unless otherwise specified, any reference to a light curve will mean a flux versus time graph.
Absolute light curves will specifically refer to absolute fluxes with times taken in the reference frame of the quasar, $t_{qso}$, while apparent light curves will refer to apparent fluxes and the times taken in the reference frame of the observer, $t_{obs}$.

On average the MACHO quasars are sampled with more regularity in the v-band.
We therefore commit our analysis only to the well sampled v-band, but in principle the finding method is independent of wavelength and may very well show less dispersion in another well sampled band.
It should also be noted that the MACHO quasars were selected based off of their variability so the below arguments may hold only for highly variable quasars.
Quasars from other surveys may clarify this.

%\\\\\\\\\\\\\\\\\\\\\\\\\\\\\\\\\\\\\\\\\\\\\\\\\\\\\\\\\\\\\\\\\\\\\\\\\\\\\
%/////////////////////////////////////////////////////////////////////////////

\section{Finding Method \& Analysis}\label{sec:findingmethod}
Our intention is to simply measure the average absolute slope of the increases and decreases in flux making up the quasar variability.
The previous work of ref.~\cite{Dai:2012wp} had hand selected approximately linear segments spanning $\sim$90 days in the quasars' rest frames and made linear fits to each.
While this was sufficient for their purposes we require a less subjective analysis.
A popular idea is to apply a neural network which is trained to select linear behaviors from noisy signals.
But as this would also require a study of the effectiveness of the neural network itself, we have decided to go with a more elementary approach.

\begin{figure}[tbp]
    \centering
    \includegraphics[width = 8.6cm]{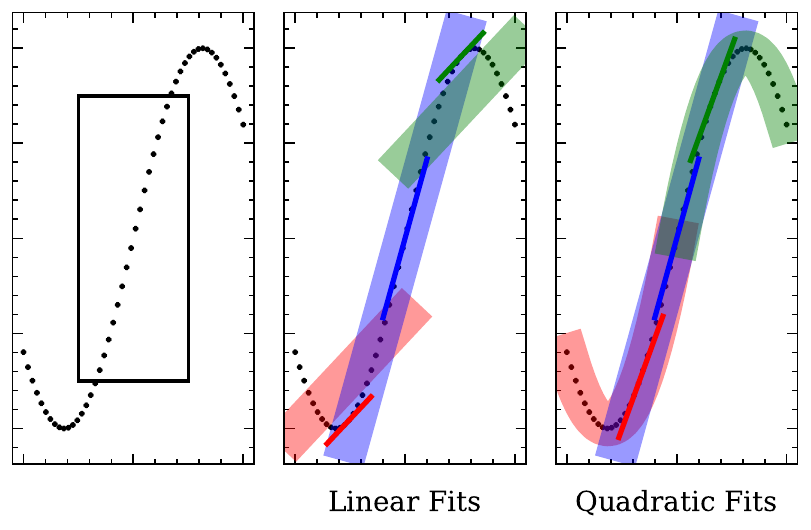}
    \caption{Illustration of the finding method using linear (center) and quadratic (right) fits. The intended slope is boxed in the leftmost figure. The thick bands in the right two figures represent fits to subsections of the curve while the lines of the same color are the slopes that are being taken from the fits. The blue line best approximates the intended slope.}
    \label{fig:linear_vs_quadratic}
\end{figure}

Searching for linear trends in noisy and intrinsically variable data is a tricky task.
It is likely that two individuals would agree on the general location of a linear segment in a varying data set but not so much on the beginning and endpoints of the linear segment.
Take figure~\ref{fig:linear_vs_quadratic} for instance where we represent a light curve with a simple sine wave shown in black dots\footnote{\label{small_trunc}Keep in mind as we go through these arguments that the actual data will have noise and a less defined pattern such that some very obvious and simple alternative solutions can become quite complicated.}.
Our interest is in the slope of the rising (and falling) segments of the curve which we have highlighted in the leftmost plot of the figure with a box.
To measure the slope of this segment one would naturally truncate the beginning and end of the curve until a roughly linear segment is obtained and then fit to a straight line.
If the truncation is not enough, resulting in contributions from data near the turning points of the curve, then the slope is skewed off of the expected value (in this case to lower values).
See footnote \ref{small_trunc} for issues with truncating too much.
In order to side step these issues we have segmented the data into subsections.
The right two plots of figure~\ref{fig:linear_vs_quadratic} shows segments spanning half the length of the total curve: the faint red bands encompass the first half of the curve, the faint green bands encompass the last half of the curve, and the faint blue bands cover the middle half of the curve, partially overlapping the previous two halves.
Further segmentation of the curve is done with five segments each spanning a third of the curve's full length and again with seven segments each spanning a fourth of the full length and so on until we reach a minimum window length which we set by hand.
We only show the first iteration for clarity.

In the linear fits plot of figure~\ref{fig:linear_vs_quadratic} the three faint bands are the linear best fits for their corresponding segments while the darker lines of the same color are the slopes of the corresponding fit at the middle point of the segment.
For the linear fit method, these two are degenerate with an overall translation.
The blue segment does pretty good and gives us an approximate value to the slope we had intended, although it is a little underestimated due to the points at its extremes.
The $\chi^2/$DoF of the blue fit would therefore be close to unity.
The red and green segments on the other hand significantly underestimate the intended slope but likewise have a poor $\chi^2/$DoF.
By taking the most common slope from the weighted population of fits, with weights determined by the $\chi^2/$DoF values, and including fits from smaller segments as discussed above we are able to get a reasonable estimate for the intended slope.
Due to the simplicity of our example all slopes will be an underestimate, but the presence of noise in actual light curves allows for overestimations as well.

The quadratic fits plot of figure~\ref{fig:linear_vs_quadratic} uses quadratic functions to fit the segments of the curve instead of linear functions as one can see from the faint bands.
The blue band is identical in both plots as expected but the red and green bands clearly fit their curve segments much better than the linear fits had.
If we evaluate the slope of the quadratic fits at the middle point of the individual segments then we will obtain the same result as what the linear fitting method gave but with worse discrimination due to better overall $\chi^2/$DoF values for each fit.
So instead we evaluate the slope of the quadratic fit at a point midway between the minimum of the fit and the maximum of the fit, hence the shifts of the red and green lines.
This gives an overall better estimate of the slope in the rising (or falling) sections of the curves as we can see from the near alignments of the red, blue, and green lines along the rising section in the curve.

While the use of linear fits has a wide spread of slope values it also has good discriminatory power through its weighting with the goodness-of-fit values.
Using quadratic fits can get a narrower spread around the intended slope but is not able to weight the slopes as well due to the overall better goodness-of-fit values of the quadratic function.
Both methods work sufficiently for the analysis but in practice the quadratic fitting method performs better.
The remainder of the article will assume the use of the quadratic method.
Also, this example only shows positive slopes but in the general case it is just the magnitude of the slope that is of interest.

One may note that this method samples the slopes of the inner regions of the curve more than the outer regions (i.e. the blue region overlaps with the inner halves of the red and green regions only).
This persists as the segmentation goes to smaller window sizes but as long as the length of the curve is much larger than the minimum window size then the effect is negligible.

We define the variational rate, $s_F$, as the most common slope returned by the above method where the subscript denotes whether the curve is from an absolute light curve, $s_F$, or an apparent light curve, $s_f$.
The method allows for an almost completely objective determination of $s_F$ with dependence on only two subjective parameters.
The first of these parameters is the lower cutoff for the sampling frequency (i.e. the average density of data points).
To get accurate fits at all timescales we require the sampling frequency to be greater than some lower cutoff value which we take to be twice per observing month.
Little dependence is seen in varying this value within reason.
Applying this cutoff avoids large unsampled gaps (usually between observing seasons) from skewing $s_F$ towards lower values.
The second parameter is a cutoff on the minimum window size the segmentation will go down to.
Considering very small segments of the light curve would allow noise to dominate over the intrinsic variations and skew our results usually towards higher values.
In order to curb this issue we have made visual inspections of the multiply lensed quasars from the COSMOGRAIL collaboration.
The overlapping regions of their matched light curves show that variations occurring at timescales greater than about 40 days are nearly identical in both images meaning that variations at timescales greater than 40 days are unlikely to be due to local noise.
So we take our minimum window cutoff to be 40 days in the quasar's rest frame.
This argument does not necessarily rule out local noise from the $>$40 day long variations in the MACHO light curves nor does it rule out interest in shorter time scales -- it just gives a standard to improve from.

\begin{figure}[tbp]
    \centering
    \includegraphics[width = 12.9cm]{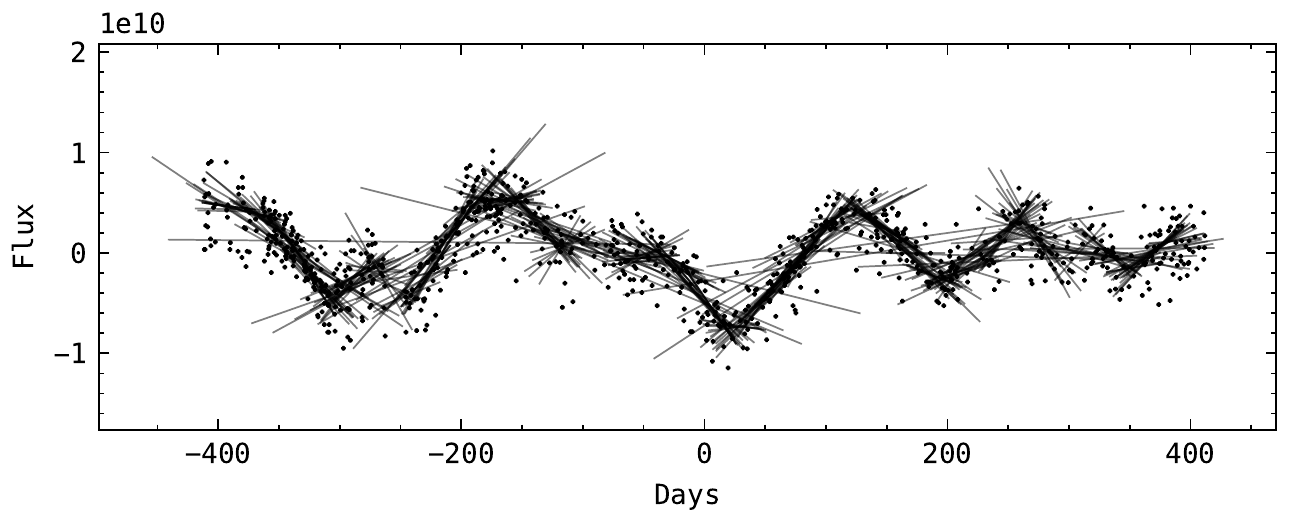}
    \caption{The fitted light curve of quasar 9.5484.258 from the MACHO project (centered at the origin). The slopes (measured using the quadratic fits method) are shown overlaying the sections of the light curve they correspond to. The fits range from half of the observation period down to $\approx 40$ days. The error bars on the data is suppressed for clarity.}
    \label{fig:fittedLC}
\end{figure}

\begin{figure}[tbp]
    \centering
    \includegraphics[width = 12.9cm]{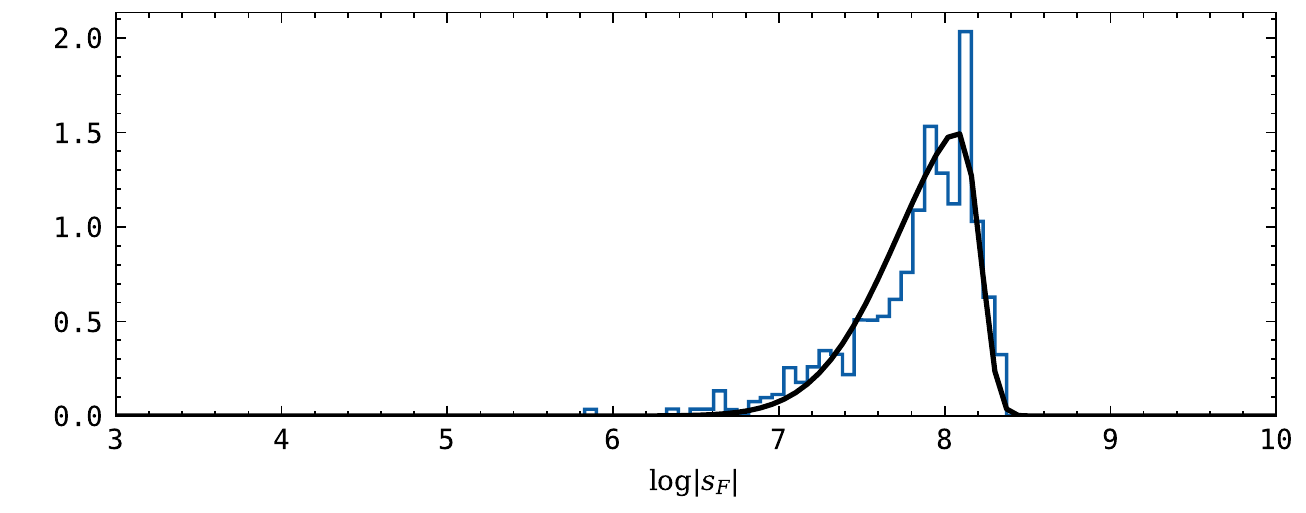}
    \caption{The population of $\log|s_F|$ for quasar 9.5484.258 weighted by the $\chi^2/$DoF of the corresponding fits. For this particular quasar, the expected variational rate would be $\log|s_F| = 8.09$.}
    \label{fig:logs_hist}
\end{figure}

\begin{figure}[tbp]
    \centering
    \includegraphics[width = \linewidth]{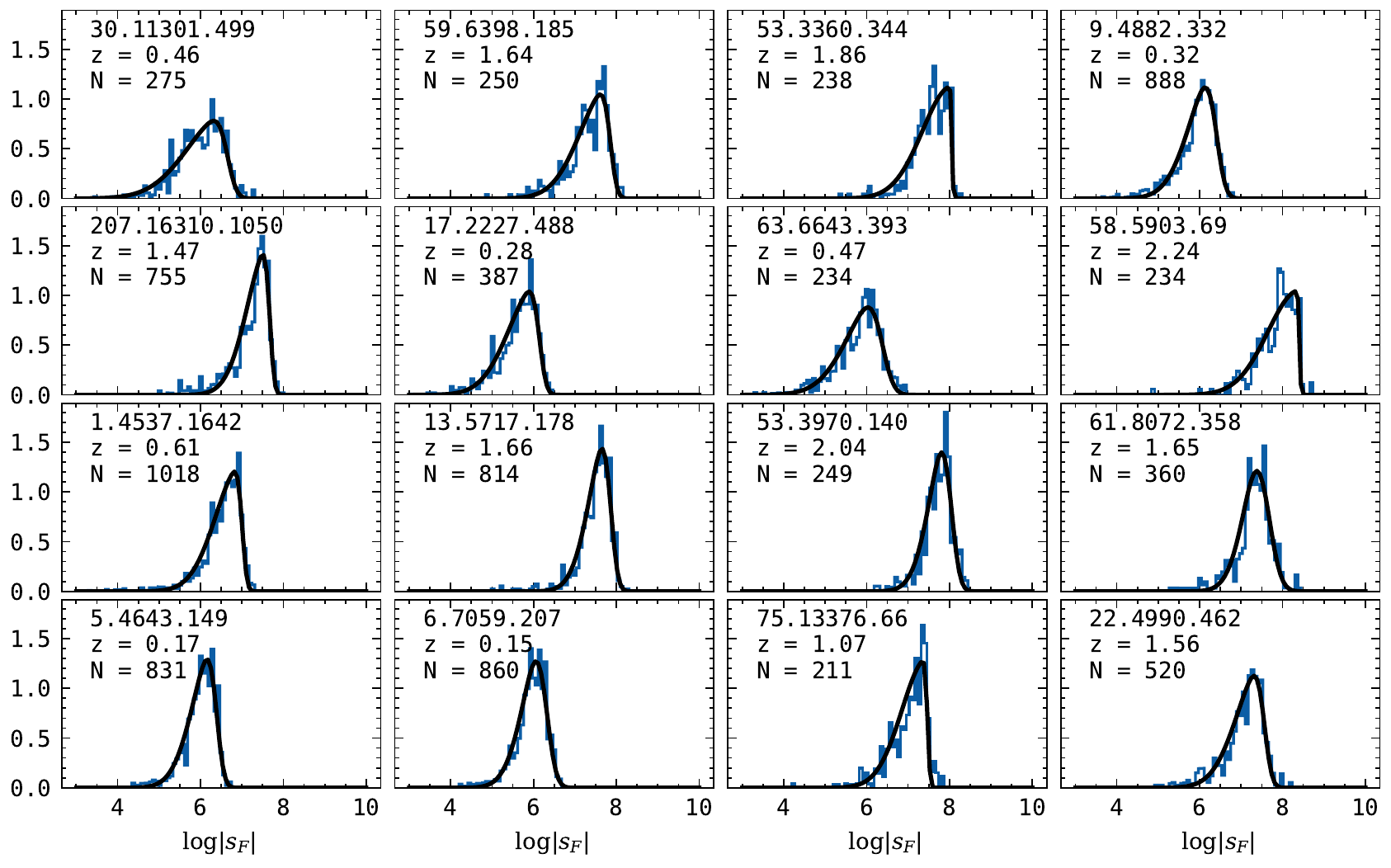}
    \caption{A larger display of the weighted $s_F$ populations in the MACHO data set. The MACHO ID, redshift, and number of observations are given for each quasar.}
    \label{fig:macho_hists}
\end{figure}

Figure~\ref{fig:fittedLC} shows the light curve of quasar 9.5484.258 with the finding method applied to it.
The slopes from the quadratic fits method are superimposed over the light curve showing a dense network of lines which map out the general behavior of the light curve.
Figure~\ref{fig:logs_hist} shows the corresponding weighted population of measured slopes which for 9.5484.258 has $\log|s_F|=8.09$ or $|s_F| = 1.23\times10^8$ units/day as determined by the mode of a skewed normal fit to the histogram (figure~\ref{fig:macho_hists} shows the weighted population of 16 other randomly selected quasars for reference).

\begin{figure}[tbp]
    \centering
    \includegraphics[width = 10.32cm]{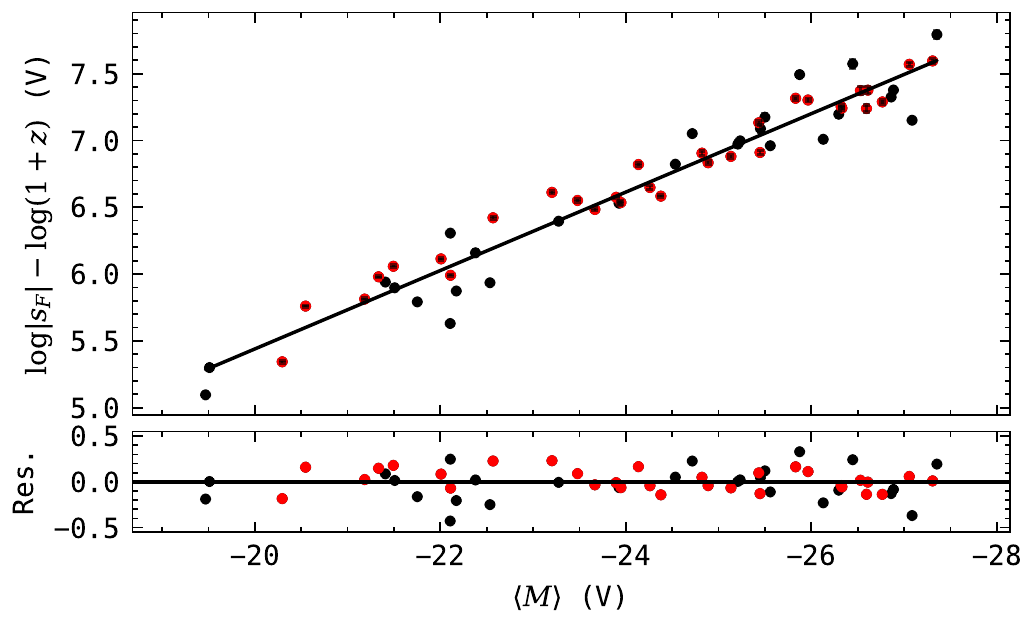}
    \caption{The v-band variational rate against the mean absolute magnitude in the quasars' rest frames shows a strong correlation with 0.15dex dispersion. Red points correspond to quasars with high sampling (N$\ge$500). Error bars are barely visible and are determined by the uncertainty in the peak position of the quasar's fitted histogram.}
    \label{fig:logs_logF}
\end{figure}

\begin{figure}[tbp]
    \centering
    \includegraphics[width = 12.9cm]{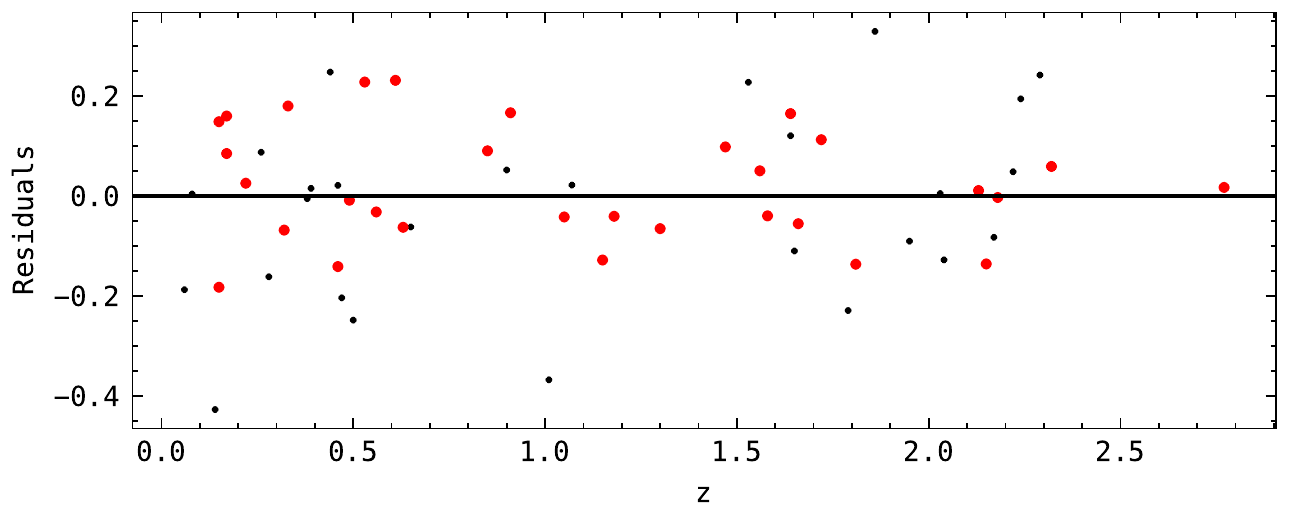}
    \caption{Residuals of the $s_F$ vs. $\langle M \rangle$ show no clear dependence on redshift.}
    \label{fig:residuals_z}
\end{figure}

Applying this analysis to each of the quasars individually we have found (see figure~\ref{fig:logs_logF}) that the variational rate has a nonlinear relation with the mean absolute magnitude,
\begin{equation}\label{eq:lineartrend}
    \log|s_F| - \log(1+z) = \alpha \langle M \rangle + \beta
\end{equation}
where the $\log(1+z)$ term is introduced to cancel out the $z$ dependence of the time derivatives in
\begin{equation*}
    s_F = \tfrac{dF}{dt_{qso}} = (1+z)\tfrac{dF}{dt_{obs}}.
\end{equation*}
Using a least squared fit gives $\alpha = -0.2929\pm0.0098$ and $\beta = -0.41\pm0.24$.
With no selection criteria made except for the two parameters already discussed we obtain a dispersion of 0.15dex without any significant dependence on $z$ as can be seen in figure~\ref{fig:residuals_z}.
If we make a very rough cut by considering only those light curves with $\ge$500 observations over the $\sim$3000 day observational period then we can reduce our dispersion down to 0.11dex though we also reduce the number of light curves to 31.
Those 31 quasars are highlighted in red in relevant figures.
We note however the bruteness of this cut since it does not take into account the uniformity of the $\ge$500 observations.
Furthermore, the model assumption used to calculate the MACHO quasars' $D_L$ values may contribute towards a higher dispersion.
It is possible that relieving the model dependence could reduce the dispersion.

For clarity, the dispersion estimates discussed here have been calculated using \eqref{eq:dispersion} where fit$\big(\langle M \rangle\big)$ refers to the right-hand side of \eqref{eq:lineartrend}.
\begin{equation}\label{eq:dispersion}
    d = \sqrt{\dfrac{\sum_i^N \Big(\text{fit}\big(\langle M \rangle_i\big) - (\log|s_F|_i - \log(1+z_i))\Big)^2}{N-1}}.
\end{equation}

%\\\\\\\\\\\\\\\\\\\\\\\\\\\\\\\\\\\\\\\\\\\\\\\\\\\\\\\\\\\\\\\\\\\\\\\\\\\\\
%/////////////////////////////////////////////////////////////////////////////

\section{Luminosity Distance}\label{sec:dl}
We now derive the luminosity distance from our fitted $s_F$ vs. $\langle M \rangle$ relation, \eqref{eq:lineartrend}.
The variational rate of the absolute light curve, $s_F$, can be related to that in the apparent light curve, $s_f = \tfrac{df}{dt_{obs}}$, through
\begin{equation}\label{eq:logsF}
    \log|s_F| = \log|s_f| + \log(1+z) + 2(\log D_L - 1) + \tfrac{2}{5}K(z).
\end{equation}
Using \eqref{eq:abs_M} and \eqref{eq:logsF} in \eqref{eq:lineartrend} allows one to obtain
\begin{equation}\label{eq:logdl}
    \log D_L = (2+5\alpha)^{-1}[\alpha\langle m \rangle + \beta  - \log|s_f|] - \tfrac{1}{5}K(z) + 1
\end{equation}
showing that in practice one can calculate the luminosity distance from the apparent light curves.

\begin{figure}[tbp]
    \centering
    \subfigure[]{
        \centering
        \includegraphics[width = 7.1cm]{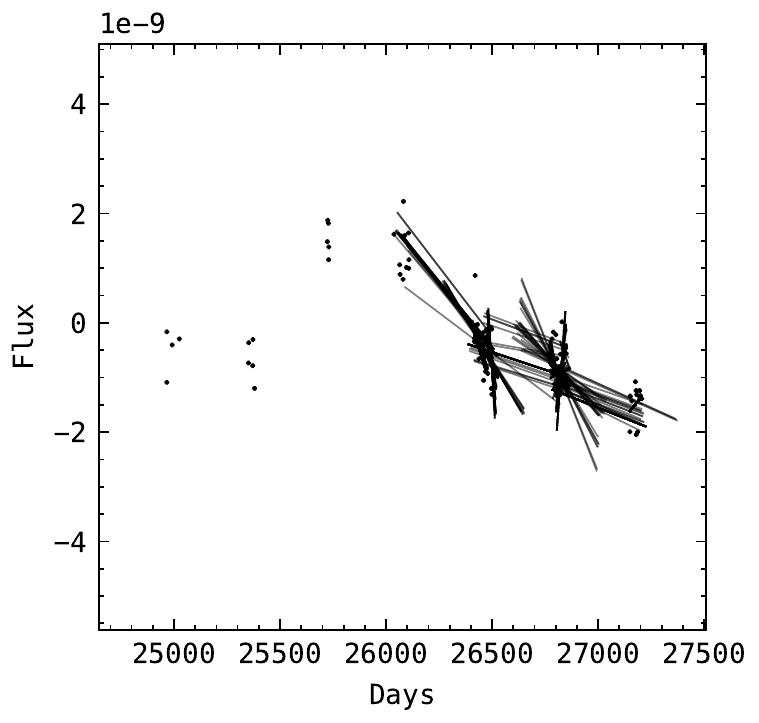}}
    \qquad
    \subfigure[]{
        \centering
        \includegraphics[width = 7.1cm]{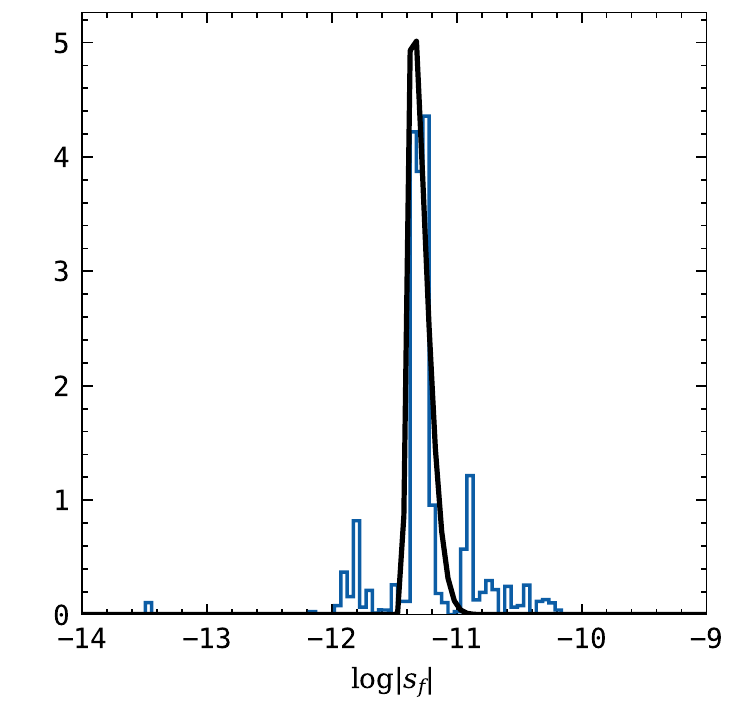}}
    \caption{(a) The fitted light curve of quasar 1365809 from the SDSS dataset (the best sampled quasar in the set). (b) The corresponding weighted $s_f$ population. To ensure the same dynamics are being compared for all quasars in the observer's frame, the minimum window size of the finding method is taken to be $(1+z)\times$40days.}
    \label{fig:sdss_lc_&_hist}
\end{figure}

In a demonstration of the work to come we have constructed a Hubble diagram for a sample of quasars from the Sloan Digital Sky Survey (SDSS) assembled by MacLeod et. al., ref.~\cite{MacLeod:2012}\footnote{We have only made use of the southern sample which can be found here: \url{http://faculty.washington.edu/ivezic/macleod/qso\_dr7/index.html}}.
That is to say, we will now calculate $\langle m \rangle$ and $s_f$ from a different set of quasar light curves and predict a cosmology from them.
The full sample contains 9258 light curves all of which have a sampling frequency far below the minimum requirements we have discussed in previous sections.
In addition, the SDSS observations were made in the u-, g-, r-, i-, and z-bands as opposed to the v-band which we have used to calibrate the $s_F$ vs. $\langle M \rangle$ relation.
But, in order to show what future work is needed, we have imposed a rough cut of $>$100 observations over the survey's $\sim$10 year observing period.
This reduces the number of light curves to 304 but allows the fitting algorithms to be applied without technical issues.
Figure~\ref{fig:sdss_lc_&_hist} show the fitted apparent light curve and resulting weighted $s_f$ population of the best sampled quasar of the SDSS dataset.
Notice the poor sampling of the quasar compared to the MACHO quasars and also the difference in time scales.

\begin{figure}[tbp]
    \centering
    \includegraphics[width = 12.9cm]{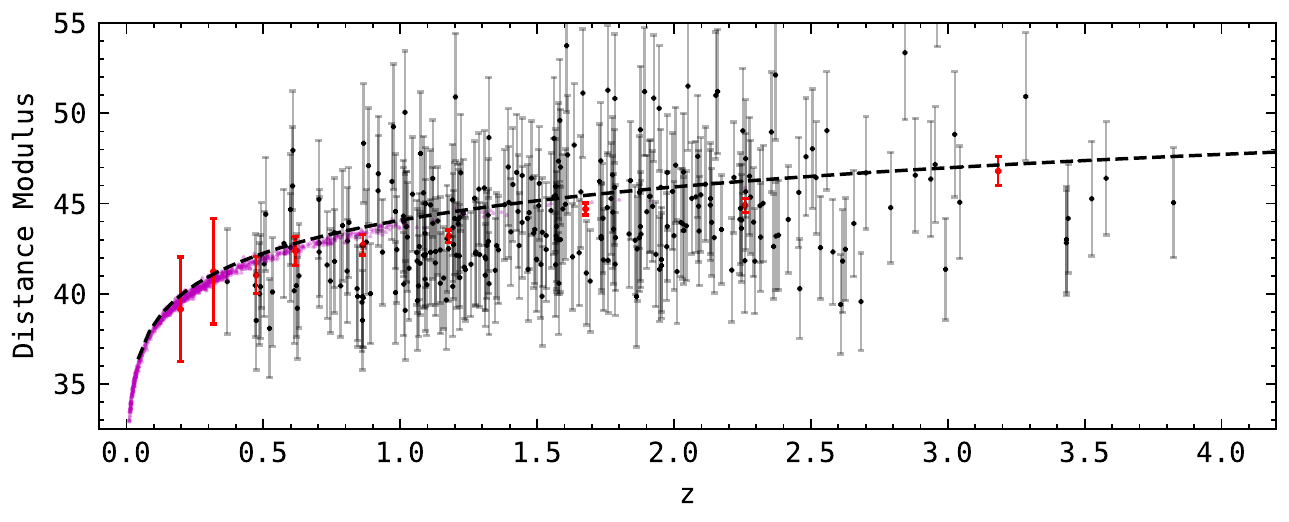}
    \caption{Hubble diagram of the 304 quasars from the SDSS sample (black points) and their binned values (red points, binned by redshift). SNe from the Pantheon survey, ref.~\cite{Scolnic:2017caz}, (magenta points) have been added for comparison. The black dashed line corresponds to the $\Lambda$CDM distance modulus \textit{Planck} 2018 best fit parameters. The quasars closely follow the $\Lambda$CDM model, accurately demonstrating our current model dependence.}
    \label{fig:hubble_diagram}
\end{figure}

Figure~\ref{fig:hubble_diagram} shows the Hubble diagram corresponding to the i-band SDSS sample with the distance modulus, DM, defined as
\begin{equation}
    DM = 5(\log D_L - 1)
\end{equation}
and errors given as
\begin{equation}
    d(DM) = 5(2+5\alpha)^{-2}\sqrt{(2\langle m \rangle - \beta + \log|s_f|)^2 d\alpha^2 + (2+5\alpha)^2(d\beta^2 + d(\log|s_f|)^2)}.
\end{equation}
The uncertainty in $\langle m \rangle$ is neglected here.
The black points correspond to the individual quasars which have been binned by redshift shown in red.
We note that these distance moduli are calculated without any assumptions on the cosmology with the exception of the model dependencies used to calibrate $\alpha$ and $\beta$.
It should also be noted that even though the SDSS light curves are significantly under sampled, the individual light curves typically have short, densely sampled observing periods which are sufficient for our finding method.
The short observing periods provide us with a rough estimate of the average slope but lack the statistics that longer observations can produce, hence the large dispersion.
Sufficiently sampled light curves should give the same overall behavior with less dispersion.

We wish to highlight that even with the exceedingly poor sampling of the SDSS dataset the general behavior still follows the $\Lambda$CDM model with good accuracy.
This should not be misinterpreted as a demonstration that $\Lambda$CDM holds up to $z\approx4$ nor should it be taken as a disagreement with refs.~\cite{Risaliti:2016nqt,Lusso:2017hgz,Salvestrini:2019thn,Lusso:2020pdb}.
Since the fitting parameters $\alpha$ and $\beta$ were calibrated using a $\Lambda$CDM cosmology, our agreement with $\Lambda$CDM only shows the effectiveness of our analysis.
Assuming a different model for the MACHO calibration would have resulted in different fitting parameters and the SDSS Hubble diagram following the distance modulus determined by that model.
Given a large enough sample of light curves, it should be possible to use the low redshift quasars in the sample to calibrate $\alpha$ and $\beta$ with minimal model dependence while using the high redshift quasars for cosmological parameter estimations.
But obtaining relative distances to well sampled quasars through their proximity to SNe would be the ideal calibration of $\alpha$ and $\beta$.
So far no such relative distances have been noted.

%\\\\\\\\\\\\\\\\\\\\\\\\\\\\\\\\\\\\\\\\\\\\\\\\\\\\\\\\\\\\\\\\\\\\\\\\\\\\\
%/////////////////////////////////////////////////////////////////////////////

\section{Damped Random Walk}\label{sec:drw}
We now comment on our agreements with findings made using synthetic quasar light curves.
Damped random walks (DRWs), or biased random walks, have proven effective towards modeling quasar light curves in the optical band (see refs.~\cite{Kelly:2009,Macleod:2010}).
DRWs perform a random walk over short time scales, $\Delta t \ll \tau$, but have a restoring term such that the walk leads back to the same place over larger time scales, $\Delta t \sim \tau$.
The process, excluding the overall magnitude, takes just two parameters: $\tau$, the relaxation time which controls the strength of the restoring term, and $SF_\infty$, the structure function at infinite time scales (not to be confused with $s_F$) which controls the variability at times $\Delta t \ll \tau$.
A DRW can be generated as a solution to the differential equation
\begin{equation}
    \tau dF(t) = -F(t)dt + SF_\infty \epsilon(t) \sqrt{2 \tau dt} + \langle F \rangle dt
\end{equation}
where $F(t)$ is the flux in our example and $\epsilon(t)$ is a white noise function with zero mean and unit variance.
We instead resort to the Python module astroML (ref.~\cite{astroML:2012}) which contains a DRW generating function built in.

We have generated 100 DRW light curves spanning an observation period of three years each.
Applying the same analysis discussed in section~\ref{sec:findingmethod} we find for the $s_F$ vs. $\langle M \rangle$ relation in the DRW light curves an $\alpha_{DRW} \approx -\tfrac{2}{5}$ with negligible deviations as we vary $\tau$ and $SF_\infty$.
No constant values of $\tau$ and $SF_\infty$ were found that would allow $\alpha_{DRW} \approx -0.3$ as one would hope in order to have agreement with the MACHO light curves.

\begin{figure}[tbp]
    \centering
    \includegraphics[width = 10.32cm]{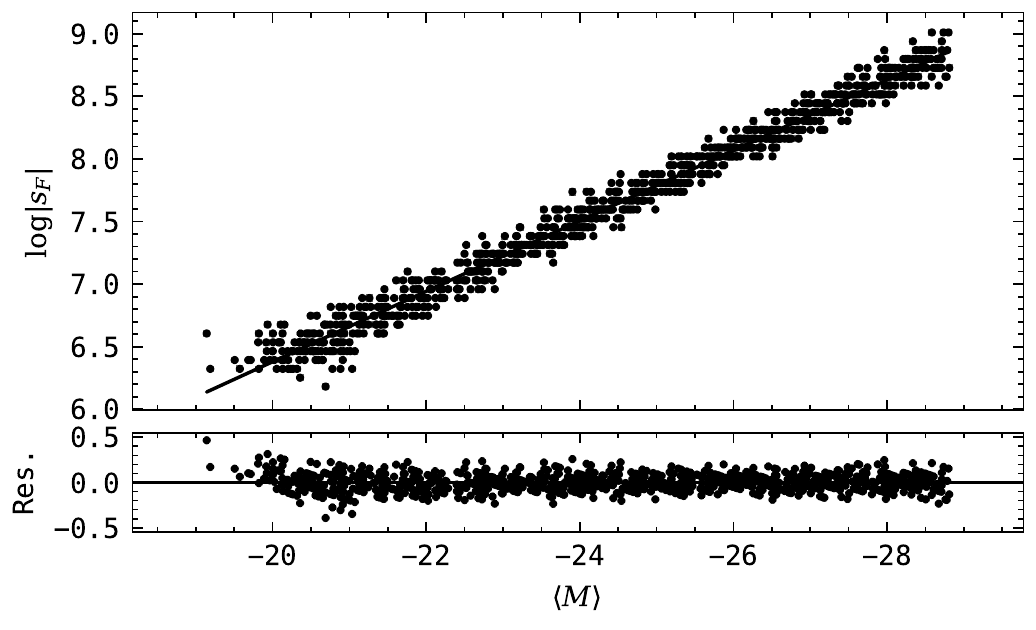}
    \caption{Same as figure~\ref{fig:logs_logF} but over the 1000 synthetic DRW light curves. For simplicity the synthetic quasars were assumed at a redshift of $z=0$.}
    \label{fig:logs_vs_logF_DRW}
\end{figure}

However, in studying the r- and b-band light curves of 100 actual quasars (55 of which were from the MACHO survey) in the context of DRWs the authors of refs.~\cite{Kelly:2009,Macleod:2010} have found relations between the mean luminosity and separately  $SF_\infty$ and $\tau$.
Other relations were also discussed but are not of interest here.
The authors of ref.~\cite{Macleod:2010} have parameterized these relations as
\begin{equation}\label{eq:SF_L}
    \log SF_\infty = A + C(M_i + 23)
\end{equation}
and
\begin{equation}\label{eq:tau_L}
    \log \tau = A' + C'(M_i + 23)
\end{equation}
where $M_i$ is the i-band mean magnitude and $\{A,C,A',C'\}$ are best fit parameters given in Table 1 of ref.~\cite{Macleod:2010}.
Logarithmic dependencies on wavelength and black hole mass have been excluded for the present work.
Now generating 1000 DRW light curves, using \eqref{eq:SF_L} and \eqref{eq:tau_L} our analysis finds an $s_F$ vs. $\langle M \rangle$ relation with $\alpha_{DRW} = -0.2813\pm0.0012$ and a dispersion of 0.095dex (see figure~\ref{fig:logs_vs_logF_DRW}).
Assuming deviations due to varying band usage to be small, it would seem as though our $\alpha = -0.2929\pm0.0098$ from the MACHO light curves is in agreement with the findings of ref.~\cite{Macleod:2010}.

Furthermore, in section~\ref{sec:dl} the luminosity distance was derived in terms of our fitting parameters from which we see that an $\alpha = -\tfrac{2}{5}$ value would not allow the determination of $D_L$ from the apparent light curves.
It would then seem that the unexpected relations between ($SF_\infty$, $\tau$) and mean luminosity that enables us to determine the luminosity distance.

One point of interest is that the use of the $SF_\infty(M_i)$ and $\tau(M_i)$ relations causes a slightly higher dispersion for low luminosity quasars (below $M>-20$ in the simulations of figure~\ref{fig:logs_vs_logF_DRW}).
It is unclear if this is a feature of quasars or an artifact of the fitted $SF_\infty(M_i)$ and $\tau(M_i)$ relations as the range of luminosities it is outside of that probed by the MACHO sample.

%\\\\\\\\\\\\\\\\\\\\\\\\\\\\\\\\\\\\\\\\\\\\\\\\\\\\\\\\\\\\\\\\\\\\\\\\\\\\\
%/////////////////////////////////////////////////////////////////////////////

\section{Conclusion}\label{sec:conclusion}
We have shown that the variability in the apparent light curves of quasars can be used to determine their luminosity distance making it possible to join quasars to the cosmic distance ladder.
This is through an observed dispersion of 0.15dex in the $s_F$ vs. $\langle M \rangle$ relation from quasars spanning a large range of luminosities.
With further studies and cleaner data sets we may be able to reduce the dispersion even further.
That being said, if the modeling of quasar light curves through DRWs is effective for all intents and purposes, the dispersion of 0.095dex in figure~\ref{fig:logs_vs_logF_DRW} may be a lower cutoff to the accuracy of this method.

Figure~\ref{fig:logs_logF} and the resulting fit, \eqref{eq:lineartrend}, show that the variations are systematically more rapid for brighter quasars in agreement with ref.~\cite{Kelly:2009,Macleod:2010}.
This would seem to disfavor the starburst model as the cause of the variations.
The starburst model, in assuming the variations are due to supernovae in the environment of the quasar, would not naturally explain the increasing variational rate of brighter quasars, but perhaps the accretion instability model could still work.
As discussed in ref.~\cite{Kelly:2009}, the short time scale variations in the optical bands are likely due to local irregularities in the accretion disk from turbulence and other effects.

The form of \eqref{eq:logdl} is similar to equation (7) in ref.~\cite{Lusso:2020pdb}.
We mention this only to clarify the advantages and limitations of our method.
Since $\log D_L$ goes as $(2+5\alpha)^{-1}$ or $0.5(1-\gamma)^{-1}$ in ref.~\cite{Lusso:2020pdb} then for our method to be on par with previous works we would need a dispersion below the $\tfrac{(2+5\alpha)}{2(1-\gamma)}$0.2dex level.
That is to say, taking $\alpha \approx -0.3$ and $\gamma \approx 0.6$, we should have a dispersion of 0.13dex or lower if we wish to improve the use of quasars as standard candles.
Using all 58 MACHO quasars does not quite satisfy this with a dispersion of 0.15dex.
Just using the better sampled quasars (N$\ge$500) does however with a dispersion of 0.11dex
The obvious drawback to our method in comparison to that in ref.~\cite{Lusso:2020pdb} is the observational effort required to produce the needed light curves.
But there have been many such observations already made by the microlensing and strong lensing communities that could relieve this issue.
A dedicated survey would however be ideal in reducing the dispersion.

At this point we have shown that our method can be applied with great success.
In follow up efforts we will extend the analyses to a larger number of low-$z$ quasar light curves in order to ensure the statistical relevance of the $s_F$ vs. $\langle M \rangle$ relation and to relieve our current model dependence.

%\\\\\\\\\\\\\\\\\\\\\\\\\\\\\\\\\\\\\\\\\\\\\\\\\\\\\\\\\\\\\\\\\\\\\\\\\\\\\
%/////////////////////////////////////////////////////////////////////////////

\acknowledgments
R.S. would like to thank Garvita Agarwal for her suggestions in the finding methods, Martin Millon for discussions on the COSMOGRAIL data, and Eoin Colg\'{a}in for discussions on the final manuscript. D.S. is partially supported by the US National Science Foundation, under Grant No. PHY-2014021.

The finding method discussed in section~\ref{sec:findingmethod} employs tools from the numpy, scipy, and matplotlib libraries, \cite{Virtanen:2019joe,Harris:2020xlr,Hunter:2007ouj}.

This paper utilizes public domain data obtained by the MACHO Project, jointly funded by the US Department of Energy through the University of California, Lawrence Livermore National Laboratory under contract No. W-7405-Eng-48, by the National Science Foundation through the Center for Particle Astrophysics of the University of California under cooperative agreement AST-8809616, and by the Mount Stromlo and Siding Spring Observatory, part of the Australian National University.

Funding for the SDSS and SDSS-II has been provided by the Alfred P. Sloan Foundation, the Participating Institutions, the National Science Foundation, the U.S. Department of Energy, the National Aeronautics and Space Administration, the Japanese Monbukagakusho, the Max Planck Society, and the Higher Education Funding Council for England. The SDSS Web Site is http://www.sdss.org/.

The SDSS is managed by the Astrophysical Research Consortium for the Participating Institutions. The Participating Institutions are the American Museum of Natural History, Astrophysical Institute Potsdam, University of Basel, University of Cambridge, Case Western Reserve University, University of Chicago, Drexel University, Fermilab, the Institute for Advanced Study, the Japan Participation Group, Johns Hopkins University, the Joint Institute for Nuclear Astrophysics, the Kavli Institute for Particle Astrophysics and Cosmology, the Korean Scientist Group, the Chinese Academy of Sciences (LAMOST), Los Alamos National Laboratory, the Max-Planck-Institute for Astronomy (MPIA), the Max-Planck-Institute for Astrophysics (MPA), New Mexico State University, Ohio State University, University of Pittsburgh, University of Portsmouth, Princeton University, the United States Naval Observatory, and the University of Washington.

\end{document}